\documentclass[aps,twocolumn,amssymb,amsmath,superscriptaddress]{revtex4-1}

\usepackage{bm}
\usepackage{color}
\usepackage{graphicx}
\usepackage{dcolumn}
\usepackage{graphicx}
\usepackage[colorlinks=true, letterpaper=true, pdfstartview=FitV, linkcolor=blue, citecolor=blue, urlcolor=blue]{hyperref}
\usepackage[normalem]{ulem}
\usepackage{amsmath}
\usepackage{epstopdf}
\usepackage{multirow}

\setcitestyle{square}

\makeatletter
\renewcommand\@biblabel[1]{#1.}
\makeatother

\UseRawInputEncoding
\begin{document}

\title{Second-order topological magneto-optical effects in noncoplanar antiferromagnets}
\author{Ping Yang}
\affiliation{Centre for Quantum Physics, Key Laboratory of Advanced Optoelectronic Quantum Architecture and Measurement (MOE), School of Physics, Beijing Institute of Technology, Beijing, 100081, China}
\affiliation{Beijing Key Lab of Nanophotonics $\&$ Ultrafine Optoelectronic Systems, School of Physics, Beijing Institute of Technology, Beijing, 100081, China}

\author{Wanxiang Feng}
\email{wxfeng@bit.edu.cn}
\affiliation{Centre for Quantum Physics, Key Laboratory of Advanced Optoelectronic Quantum Architecture and Measurement (MOE), School of Physics, Beijing Institute of Technology, Beijing, 100081, China}
\affiliation{Beijing Key Lab of Nanophotonics $\&$ Ultrafine Optoelectronic Systems, School of Physics, Beijing Institute of Technology, Beijing, 100081, China}

\author{Xiaodong Zhou}
\affiliation{Centre for Quantum Physics, Key Laboratory of Advanced Optoelectronic Quantum Architecture and Measurement (MOE), School of Physics, Beijing Institute of Technology, Beijing, 100081, China}
\affiliation{Beijing Key Lab of Nanophotonics $\&$ Ultrafine Optoelectronic Systems, School of Physics, Beijing Institute of Technology, Beijing, 100081, China}
\affiliation {Laboratory of Quantum Functional Materials Design and Application, School of Physics and Electronic Engineering, Jiangsu Normal University, Xuzhou 221116, China}

\author{Xiuxian Yang}
\affiliation{Centre for Quantum Physics, Key Laboratory of Advanced Optoelectronic Quantum Architecture and Measurement (MOE), School of Physics, Beijing Institute of Technology, Beijing, 100081, China}
\affiliation{Beijing Key Lab of Nanophotonics $\&$ Ultrafine Optoelectronic Systems, School of Physics, Beijing Institute of Technology, Beijing, 100081, China}

\author{Yugui Yao}
\affiliation{Centre for Quantum Physics, Key Laboratory of Advanced Optoelectronic Quantum Architecture and Measurement (MOE), School of Physics, Beijing Institute of Technology, Beijing, 100081, China}
\affiliation{Beijing Key Lab of Nanophotonics $\&$ Ultrafine Optoelectronic Systems, School of Physics, Beijing Institute of Technology, Beijing, 100081, China}

\date{\today}

\begin{abstract}
The second-order magneto-optical effects, represented by Voigt and Sch\"{a}fer-Hubert effects, are effective methods to detect the spin textures in antiferromagnets, whereas the previous studies are usually limited to collinear antiferromagnets.  In noncollinear antiferromagnets, the spin textures characterized by spin chirality have been revealed to play a critical role in many exciting physics.  In particular, the first-order topological magneto-optical effects originated from scalar spin chirality have been discovered recently.  In this work, using the first-principles calculations and group theory analysis, we generalize the first-order topological magneto-optical effects to the second-order cases, that is, topological Voigt and Sch\"{a}fer-Hubert effects, by taking the noncoplanar 3Q spin state of $\gamma$-Fe$_{x}$Mn$_{1-x}$ alloy as an example.  The conventional Voigt and Sch\"{a}fer-Hubert effects are comparatively studied in the collinear 1Q and 2Q  spin states of $\gamma$-Fe$_{x}$Mn$_{1-x}$ alloy.  In addition, the natural linear birefringence due to crystal anisotropy is discussed in the strained 1Q, 2Q, and 3Q states, and a unique fingerprint for experimentally distinguishing the second-order topological magneto-optical effects and natural linear birefringence is identified.  Our work brings a topological insight into the second-order magneto-optical effects in noncoplanar antiferromagnets and also provides $\gamma$-Fe$_{x}$Mn$_{1-x}$ as an attractive material platform for future experimental exploration.
\end{abstract}

\maketitle

\section{Introduction} \label{sec:Introduction}

Antiferromagnets (AFMs) have promising applications in high-density spintronic devices due to their excellent properties, such as the absence of stray fields, robustness to magnetic field perturbation, and ultrafast spin dynamics~\cite{Jungwirth2016,Baltz2018}.  Except for common collinear AFMs, the noncollinear magnetic orders are also widely seen in AFMs.  The noncollinear AFMs can be further classified into coplanar and noncoplanar ones, which are characterized by vector and scalar spin chiralities~\cite{Kawamura2001},
\begin{eqnarray}
	\boldsymbol{\kappa} &=& \sum_{<ij>}\mathbf{S}_{i}\times\mathbf{S}_{j}, \label{eq:kappa}\\
	\chi &=& \sum_{<ijk>}\mathbf{S}_{i}\cdot\left(\mathbf{S}_{j}\times\mathbf{S}_{k}\right), \label{eq:chi}
\end{eqnarray}
respectively, where $\mathbf{S}_{i}$, $\mathbf{S}_{j}$, and $\mathbf{S}_{k}$ are the spins on neighboring sublattices.  Recently, various unexpected physical phenomena were reported in noncollinear chiral AFMs.  For example, the magneto-optical Kerr and Faraday effects (MOKE and MOFE), which are usually known to be linear in magnetization $\mathbf{M}$, are believed non-existent in AFMs due to their zero net magnetization.  Nevertheless, with the allowance of special magnetic symmetries, MOKE and MOFE have been theoretically predicted or experimentally observed in coplanar noncollinear AFMs Mn$_3X$ ($X$ = Rh, Ir, Pt)~\cite{WX-Feng2015}, Mn$_3Y$ ($Y$ = Ge, Ga, Sn)~\cite{Higo2018,Balk2019,MX-Wu2020}, and Mn$_3Z$N ($Z$ = Ga, Zn, Ag, Ni)~\cite{XD-Zhou2019a} as well as in noncoplanar AFMs $\gamma$-Fe$_{0.5}$Mn$_{0.5}$~\cite{WX-Feng2020} and K$_{0.5}$RhO$_{2}$~\cite{WX-Feng2020}.  The MOKE and MOFE unveiled in noncoplanar AFMs were termed \textit{topological} magneto-optical effects as they originate from scalar spin chirality instead of spin-orbit coupling and band exchange splitting, fundamentally distinguishing them from conventional magneto-optical effects~\cite{WX-Feng2020}.
	
\begin{figure}
	\includegraphics[width=1\columnwidth]{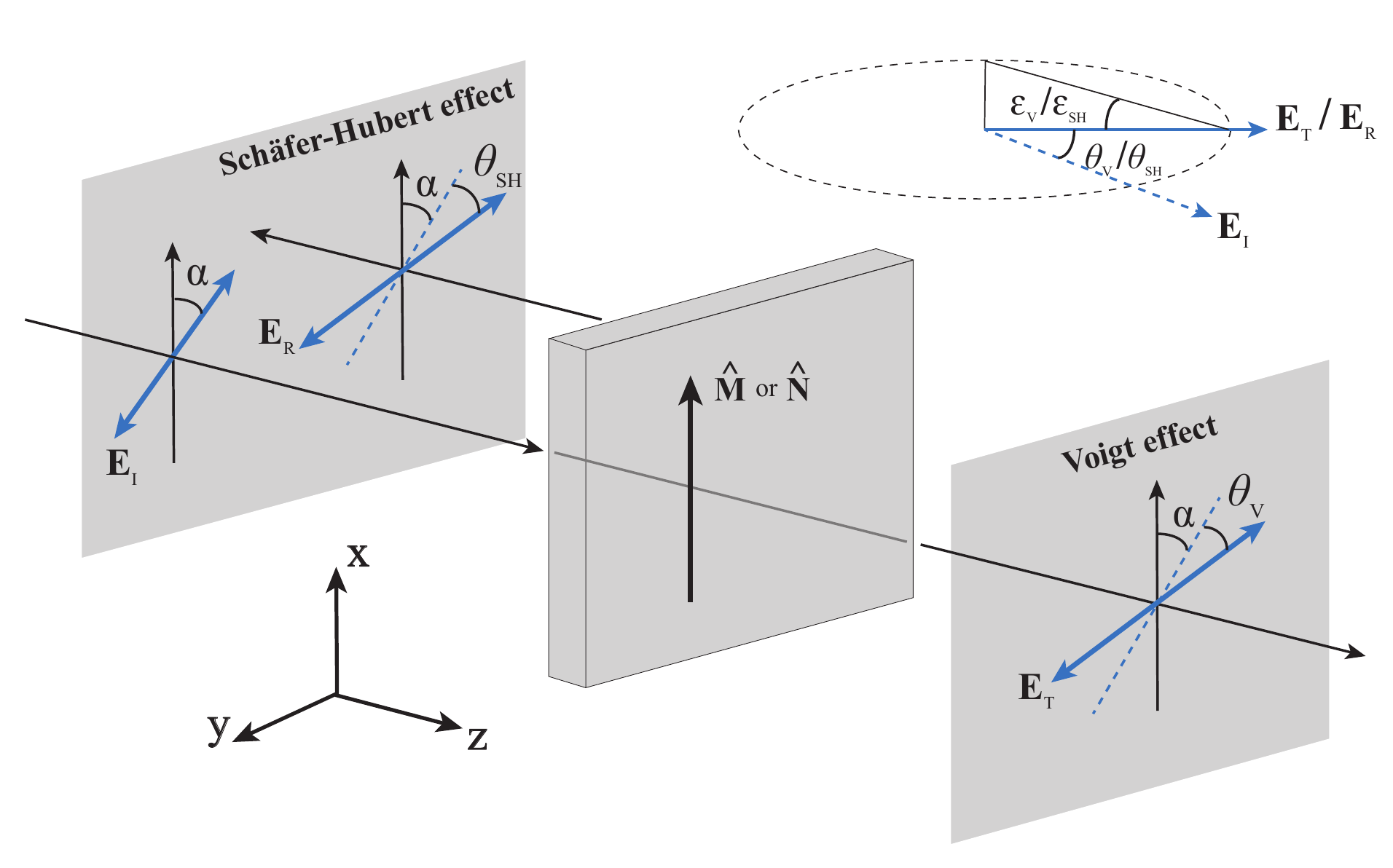}
	\caption{Schematic illustration of magneto-optical Voigt and Sch\"{a}fer-Hubert effects.  The incident linearly polarized light propagating along the $z$-axis normally shines on the surface ($xy$-plane) of an in-plane magnetized material (assuming $x$-axis).  There exists an angle ($\alpha$) between the electric field of incident light ($\mathbf{E}_\textnormal{I}$) and the magnetization direction of materials ($\hat{\mathbf{N}}$ for AFMs or $\hat{\mathbf{M}}$ for FMs).  The transmitted and reflected lights become elliptically polarized accompanying by the rotations of polarization planes.  The rotation angles, $\theta_\textnormal{V}$ and $\theta_\textnormal{SH}$, describe the deflections of $\mathbf{E}_\textnormal{T}$ and $\mathbf{E}_\textnormal{R}$ with respect to $\mathbf{E}_\textnormal{I}$.  The ellipticities, $\varepsilon_\textnormal{V}$ and $\varepsilon_\textnormal{SH}$, are the quotient of the short and long axes of the ellipse.}
	\label{fig:model}
\end{figure}

The MOKE and MOFE that appeared in AFMs with zero net magnetization can still be regarded as the linear (first-order) magneto-optical effects since they are odd in $\hat{\mathbf{N}}=\mathbf{N}/|\mathbf{N}|$, where $\mathbf{N}$ is the N\'{e}el vector.  That means, the signs of MOKE and MOFE will change if the direction of $\hat{\mathbf{N}}$ reverses.  One would naturally concern about whether the nonlinear (e.g., second-order) magneto-optical effects exist in noncollinear chiral AFMs.  Among the second-order magneto-optical effects, the Voigt~\cite{Voigt1908} and Sch\"{a}fer-Hubert~\cite{Schafer1990} effects are two representative ones, which refer to the rotation of the polarization plane of a linearly polarized light normally transmitted through and reflected from an in-plane magnetized material, respectively (Fig.~\ref{fig:model}).  It should be noted that the terminologies used for magneto-optical Voigt and Sch\"{a}fer-Hubert effects (MOVE and MOSHE) in the literature are discrepant --- some measurements in a reflection geometry are called MOVE but not MOSHE and the MOVE is sometimes denoted as Cotton-Mouton effect or magnetic linear birefringence or magnetic linear dichroism.  Very recently, the time-resolved MOVE (actually measured in a reflection geometry) has been observed in a coplanar noncollinear AFM Mn$_3$Sn using the pump-probe experimental technique, and the modulated Voigt angle is surprisingly one order of magnitude larger than the Kerr angle~\cite{HC-Zhao2021}.  However, so far, little is known about the MOVE and MOSHE in noncoplanar AFMs.

The scalar spin chirality $\chi$ is a critical quantity in noncoplanar AFMs, which brings many exciting physics.  Considering an electron moves along a close path connected by three noncoplanar spin sublattices ($\mathbf{S}_{i}$, $\mathbf{S}_{j}$, and $\mathbf{S}_{k}$), the electron shall experience a fictitious magnetic flux that is proportional to the scalar spin chirality, $\mathbf{b}_{f}\propto t_{3}\chi_{ijk}\mathbf{\hat{n}}$, where $t_{3}=t_{ij}t_{jk}t_{ki}$ is the successive transfer integral along the path ($i\rightarrow j\rightarrow k\rightarrow i$) and $\mathbf{\hat{n}}$ is a unit vector normal to the plane formed by the three sublattices.  This is the so-called real space Berry phase effect, which is responsible for the topological Hall effect~\cite{Bruno2004,Franz2014} and its quantization~\cite{J-Zhou2016}, topological orbital magnetism~\cite{Hanke2017}, and the first-order topological magneto-optical effects~\cite{WX-Feng2020} and their quantization~\cite{WX-Feng2020}.  It is reasonable to speculate that the second-order topological magneto-optical effects (such as topological MOVE and MOSHE), originating from scalar spin chirality and being even in $\hat{\mathbf{N}}$, can exist in noncoplanar AFMs as well.  In other words, the topological magneto-optical effects should be able to generalize from the first-order cases (MOKE and MOFE) to the second-order cases (MOVE and MOSHE).  The topological MOVE and MOSHE in noncoplanar AFMs are expected to differ from conventional MOVE and MOSHE in ferromagnets (FMs) ~\cite{Mertins2001,Valencia2010} and collinear AFMs~\cite{Saidl2017}, which are dependent on spin-orbit coupling and band exchange splitting.

In this work, using the first-principles calculations together with magnetic group analysis, we explore the topological MOVE and MOSHE in noncoplanar antiferromagnetic $\gamma$-Fe$_{x}$Mn$_{1-x}$ alloy.  The theory and computational methods for the second-order magneto-optical effects are addressed detailedly in Sec.~\ref{sec:Theory}.  The optical geometries for observing the MOVE and MOSHE are proposed for the collinear 1Q and 2Q spin states as well as the noncoplanar 3Q spin state.  The magnetic group theory is then applied to determine the shape of the permittivity tensor, which is a crucial step in calculating the MOVE and MOSHE.  In Sec.~\ref{sec:1Q}, the conventional MOVE and MOSHE effects are found in collinear 1Q and 2Q  states of $\gamma$-Fe$_{x}$Mn$_{1-x}$.  The calculated Voigt and Sch\"{a}fer-Hubert angles are comparable or even larger than that of the famous collinear AFM CuMnAs.  The natural linear birefringence (NLB) due to crystal anisotropy in the strained 1Q and 2Q  states are at least one order of magnitude smaller than the MOVE and MOSHE.  In Sec.~\ref{sec:3Q}, we reveal the topological MOVE and MOSHE originating from scalar spin chirality in the noncoplanar 3Q state of $\gamma$-Fe$_{x}$Mn$_{1-x}$.  Since the strain simultaneously induces scalar spin chirality and crystal anisotropy in the strained 3Q state, the topological MOVE/MOSHE and NLB are mixed together.  We find that the magnitudes of topological MOVE/MOSHE and NLB are in the same order of magnitude, but their signs are opposite.  A unique fingerprint for experimentally distinguishing the topological MOVE/MOSHE and NLB is identified.  Finally, a brief summary is given in Sec.~\ref{sec:Summary}.  Our work suggests that the antiferromagnetic $\gamma$-Fe$_{x}$Mn$_{1-x}$ alloy is an interesting material platform for exploring the novel high-order topological light-matter interactions.

\section{Theory and computational details} \label{sec:Theory}

For crystallographically-isotropic nonmagnetic materials, the application of an external magnetic field lowers the symmetry of systems, leading to a change in the shape of the permittivity tensor.  Specifically, if a linearly polarized light normally shines on the surface of materials and the magnetic field lies on the surface (i.e., normal to the propagating direction of incident light), the light passing through the materials shall have two different indices of refraction, $n_{\parallel}$ and $n_{\perp}$, which are parallel to and perpendicular to the external magnetic field, respectively.  Once the angle between the electric field of incident light and the external magnetic field does not equal 0$^\circ$ or 90$^\circ$, the transmitted and reflected lights have to become elliptically polarized, accompanying by the rotations of polarization planes.  If the external magnetic field is replaced by spontaneous magnetic orders (FMs or AFMs), the resultant magneto-optical phenomena resemble, named Voigt~\cite{Voigt1908} and Sch\"{a}fer-Hubert effects~\cite{Schafer1990} building in transmission and reflection geometries, respectively (Fig.~\ref{fig:model}).  It should be mentioned that for crystallographically-anisotropic nonmagnetic materials, the above optical phenomena still exist if the electric field of incident light is not parallel or perpendicular to the optical axis, but they are usually called NLB.  In the following, we shall discuss how to separate the NLB from the magnetism-induced Voigt and Sch\"{a}fer-Hubert effects.

For the MOVE, the rotation angle ($\theta_{\textnormal{V}}$) and ellipticity ($\varepsilon_{\textnormal{V}}$) of elliptically-polarized transmitted light are usually combined into the so-called complex Voigt angle,
\begin{equation}\label{eq:Voigt-alpha}
	\Phi_{\textnormal{V}}(\alpha)= \left(\varepsilon_{\textnormal{V}} + i \theta_{\textnormal{V}}\right) \sin(2\alpha), \\
\end{equation}	
where $\alpha$ is the angle between the electric field of incident light ($\textbf{E}_\textnormal{I}$) and the magnetization direction ($\hat{\mathbf{N}}$ for AFMs or $\hat{\mathbf{M}}=\mathbf{M}/|\mathbf{M}|$ for FMs).  It is clear that $\Phi_{\textnormal{V}}(\alpha)$ reaches its maximum when $\alpha = 45^{\circ}$, which corresponds to the standard Voigt geometry often employed in experiments.  And $\Phi_{\textnormal{V}}(\alpha)$ is certainly zero if $\alpha = 0^{\circ}$ or $90^{\circ}$.  Since the evolution of $\Phi_{\textnormal{V}}(\alpha)$ with $\alpha$ is just a sinusoidal curve with a period of $\pi$, hereafter, we shall only discuss the maximal value of $\Phi_{\textnormal{V}}(\alpha)$, which is expressed as~\cite{Mertins2001},
\begin{equation}\label{eq:Voigt}
	\Phi_{\textnormal{V}}^{\textnormal{max}} = \varepsilon_{\textnormal{V}} + i \theta_{\textnormal{V}} \approx \frac{\omega d}{2c} (n_{\parallel}-n_{\perp}),
\end{equation}
where $\omega$ is light frequency, $c$ is the speed of light in vacuum, and $d$ is the thickness of materials.  In our actual calculations, the Voigt angles are accounted in the unit of deg/cm, and therefore the thickness $d$ is not specified.  Without loss of generality, we assume that $\hat{\mathbf{N}}$ is parallel to the $x$-axis and the incident light propagates along the $z$-axis.  By solving the Fresnel equation, the refractive indices that are parallel to and perpendicular to $\hat{\mathbf{N}}$ can be written as,
\begin{eqnarray}
	n_{\parallel} &=& \sqrt{\epsilon_{xx}}, \label{eq:n_parallel} \\
	n_{\perp} &=& \sqrt{\epsilon_{yy} + \epsilon_{yz}^{2} / \epsilon_{zz}}, \label{eq:n_perp}
\end{eqnarray}
where $\epsilon_{\mu\nu}$ with $\mu,\nu \in \{x,y,z\}$ is the permittivity tensor.  We can further obtain
\begin{eqnarray}
	n_{\parallel} n_{\perp} &\approx& \frac{1}{2}(\epsilon_{xx} + \epsilon_{yy} + \epsilon_{yz}^2/\epsilon_{zz}),  \label{eq:n_pn_p} \\
	n_{\parallel} - n_{\perp} &\approx& \frac{1}{2\bar{n}}(\epsilon_{xx} - \epsilon_{yy} - \epsilon_{yz}^2/\epsilon_{zz}), \label{eq:n_p-n_p}
\end{eqnarray}
where $\bar{n} = \frac{1}{2}(n_{\parallel}+n_{\perp}) $ is the average of the refractive indices.

For the MOSHE, the rotation angle ($\theta_{\textnormal{SH}}$) and ellipticity ($\varepsilon_{\textnormal{SH}}$) of elliptically-polarized reflected light can be similarly written as the complex Sch\"{a}fer-Hubert angle,
\begin{equation}\label{eq:SH-alpha}
	\Phi_{\textnormal{SH}}(\alpha)= \left(\theta_{\textnormal{SH}} + i\varepsilon_{\textnormal{SH}}\right) \sin(2\alpha), \\
\end{equation}
which also displays a period of $\pi$ with respect to the angle $\alpha$.  The maximal value of $\Phi_{\textnormal{SH}}(\alpha)$ occurs at $\alpha=45^\circ$, that is~\cite{Valencia2010},
\begin{equation}\label{eq:SH}
	\Phi_{\textnormal{SH}}^{\textnormal{max}} = \theta_{\textnormal{SH}} + i \varepsilon_{\textnormal{SH}}  \approx  \frac{n_{\parallel}-n_{\perp}}{1 - n_{\parallel}n_{\perp}}.
\end{equation}

One can read from Eqs.~\eqref{eq:Voigt} and~\eqref{eq:SH} that the birefringence, $\Delta n= n_{\parallel}-n_{\perp}$, is the dominating quantity for both MOVE and MOSHE.  If a magnetic material presents in-plane crystal anisotropy (e.g., a $C^{z}_{2}$ symmetry lives on the $xy$-plane), a nonzero component of $\Delta n$ contributing from the NLB should be expected.  This part has to be separated from total birefringence if one wishes to find out the magnetism-induced MOVE and MOSHE, just like what have done in crystallographically-anisotropic Cd$_{1-x}$Mn$_{x}$Se~\cite{Oh1991}.  To do this, we expand the permittivity tensor into a Taylor series in powers of magnetization~\cite{Tesarova2014,Zelezny2017,XX-Yang2022},
\begin{equation}\label{eq:permittivity}
	\epsilon_{\mu\nu}(\hat{\mathbf{N}})=\epsilon^{(0)}_{\mu\nu}+\epsilon^{(1)}_{\mu\nu} \hat{\mathbf{N}}+\epsilon^{(2)}_{\mu\nu} \hat{\mathbf{N}}^{2} + \cdots,
\end{equation}
where $\epsilon^{(0)}_{\mu\nu}$ is the part independent on magnetization, $\epsilon^{(1)}_{\mu\nu}$ and $\epsilon^{(2)}_{\mu\nu}$ are linearly and quadratically dependent on magnetization, respectively.  In principle, we can calculate $\epsilon_{\mu\nu}$ for a given magnetic material and calculate $\epsilon^{(0)}_{\mu\nu}$ by artificially removing the magnetic order.  Since $\epsilon^{(0)}_{\mu\nu}$ is responsible for the NLB due to crystal anisotropy, subtracting $\epsilon^{(0)}_{\mu\nu}$ from $\epsilon_{\mu\nu}$ is the part purely originating from magnetism, being responsible for the MOVE and MOSHE.

According to the Onsager relation, $\epsilon_{\mu\nu} (\hat{\mathbf{N}})=\epsilon_{\nu\mu} (-\hat{\mathbf{N}})$, we know $\epsilon^{(1)}_{\mu\mu}=0$, $\epsilon^{(0)}_{\mu\nu} = \epsilon^{(0)}_{\nu\mu}$, $\epsilon^{(1)}_{\mu\nu} = -\epsilon^{(1)}_{\nu\mu}$, and $\epsilon^{(2)}_{\mu\nu} = \epsilon^{(2)}_{\nu\mu}$.  Omitting the magnetism-independent parts, the diagonal terms of the permittivity tensor are square of $\hat{\mathbf{N}}$, i.e., $\tilde{\epsilon}_{\mu\mu}(\hat{\mathbf{N}}) = \epsilon^{(2)}_{\mu\mu} \hat{\mathbf{N}}^{2}$, while the off-diagonal terms of the permittivity tensor have first-order antisymmetric and second-order symmetric parts, i.e., $\tilde{\epsilon}_{\mu\nu}(\hat{\mathbf{N}})=\epsilon^{(1)}_{\mu\nu} \hat{\mathbf{N}}+\epsilon^{(2)}_{\mu\nu} \hat{\mathbf{N}}^{2}$.  Since $n_{\parallel} - n_{\perp}$ and $n_{\parallel}n_{\perp}$ are linearly (quadratically) dependent on the diagonal (off-diagonal) terms of the permittivity tensor (Eqs.~\eqref{eq:n_pn_p} and~\eqref{eq:n_p-n_p}), $\Phi_{\textnormal{V}}^{\textnormal{max}}$ and $\Phi_{\textnormal{SH}}^{\textnormal{max}}$ must be even in $\hat{\mathbf{N}}$ and to the lowest-order quadratic to $\hat{\mathbf{N}}$.  It is in sharp contrast to the MOKE and MOFE, which are odd in $\hat{\mathbf{N}}$ and solely depend on $\epsilon^{(1)}_{\mu\nu}$.  In many AFMs, the MOKE and MOFE are forbidden because the off-diagonal terms of the permittivity tensor are forced to be zero under certain symmetries (e.g., the $\mathcal{PT}$ symmetry, $\mathcal{P}$ and $\mathcal{T}$ are space inversion and time-reversal operations, respectively).  Conversely, the MOVE and MOSHE can exist in AFMs due to the inequality of two diagonal terms, $\epsilon_{\mu\mu} \neq \epsilon_{\nu\nu}$, even though the off-diagonal term $\epsilon_{\mu\nu}$ is possibly zero, refer to Eq.~\eqref{eq:n_p-n_p}.  Therefore, the MOVE and MOSHE take a great advantage over the MOKE and MOFE for studying AFMs.

The key ingredient for calculating the MOVE and MOSHE is the permittivity tensor or the optical conductivity in an equivalent way, since $\epsilon_{\mu\nu}=\delta_{\mu\nu} + \frac{4\pi i}{\omega} \sigma_{\mu\nu}$, where  $\delta_{\mu\nu}$ equals 1 if $\mu =\nu$ and 0 otherwise.  The optical conductivity can be evaluated using the Kubo formula~\cite{Yates2007} implemented in \textsc{wannier90} package~\cite{Pizzi2020},
\begin{eqnarray}\label{eq:OPC}
	\sigma_{\mu\nu}&=& \frac{ie^2\hbar}{N_k V}\sum_{\textbf{k}}\sum_{n, m}\frac{f_{m\textbf{k}}-f_{n\textbf{k}}}{E_{m\textbf{k}}-E_{n\textbf{k}}} \nonumber\\
	&&\times\frac{\langle\psi_{n\textbf{k}}|\hat{\upsilon}_{\mu}|\psi_{m\textbf{k}}\rangle\langle\psi_{m\textbf{k}}|\hat{\upsilon}_{\nu}|\psi_{n\textbf{k}}\rangle}{E_{m\textbf{k}}-E_{n\textbf{k}}-(\hbar\omega+i\eta)},
\end{eqnarray}
where $f_{n\textbf{k}}$, $V$, $N_k$, $\hat{\upsilon}_{\mu(\nu)}$, $\hbar\omega$, and $\eta$ are the Fermi-Dirac distribution function, volume of the unit cell, total number of $k$-points for sampling the Brillouin zone, velocity operators, photon energy, and energy smearing parameter, respectively.  $\psi_{n\textbf{k}}$ and $E_{n\textbf{k}}$ are the Bloch wavefunction and interpolated energy at the band index $n$ and momentum $\textbf{k}$, respectively.  A $k$-points mesh of $100 \times 100 \times 100$ is enough to converge the optical conductivity, and $\eta$ is set to be 0.1 eV.  Because of the metallic nature of $\gamma$-Fe$_{x}$Mn$_{1-x}$, the intraband transitions can not be ignored in the low energy range (e.g., $<$ 1.0 eV), and the Drude term $\sigma_{\textnormal{D}} = \sigma_{0} /\left(1-i \omega \tau_{\textnormal{D}}\right)$ is added into the diagonal terms of optical conductivity.  The Drude parameters $\sigma_{0}$ and $\tau_{\textnormal{D}}$ are obtained by linearly interpolating the experimental data of pure Fe ($\sigma_{0} = 6.40\times10^{15}$ s$^{-1}$ and $\tau_{\textnormal{D}} = 9.12\times10^{-15}$ s)~\cite{Lenham1966} and pure Mn ($\sigma_{0} = 4.00\times10^{15}$ s$^{-1}$ and $\tau_{\textnormal{D}} = 0.33\times10^{-15}$ s)~\cite{Lenham1966}.

\begin{figure}
	\includegraphics[width=0.9\columnwidth]{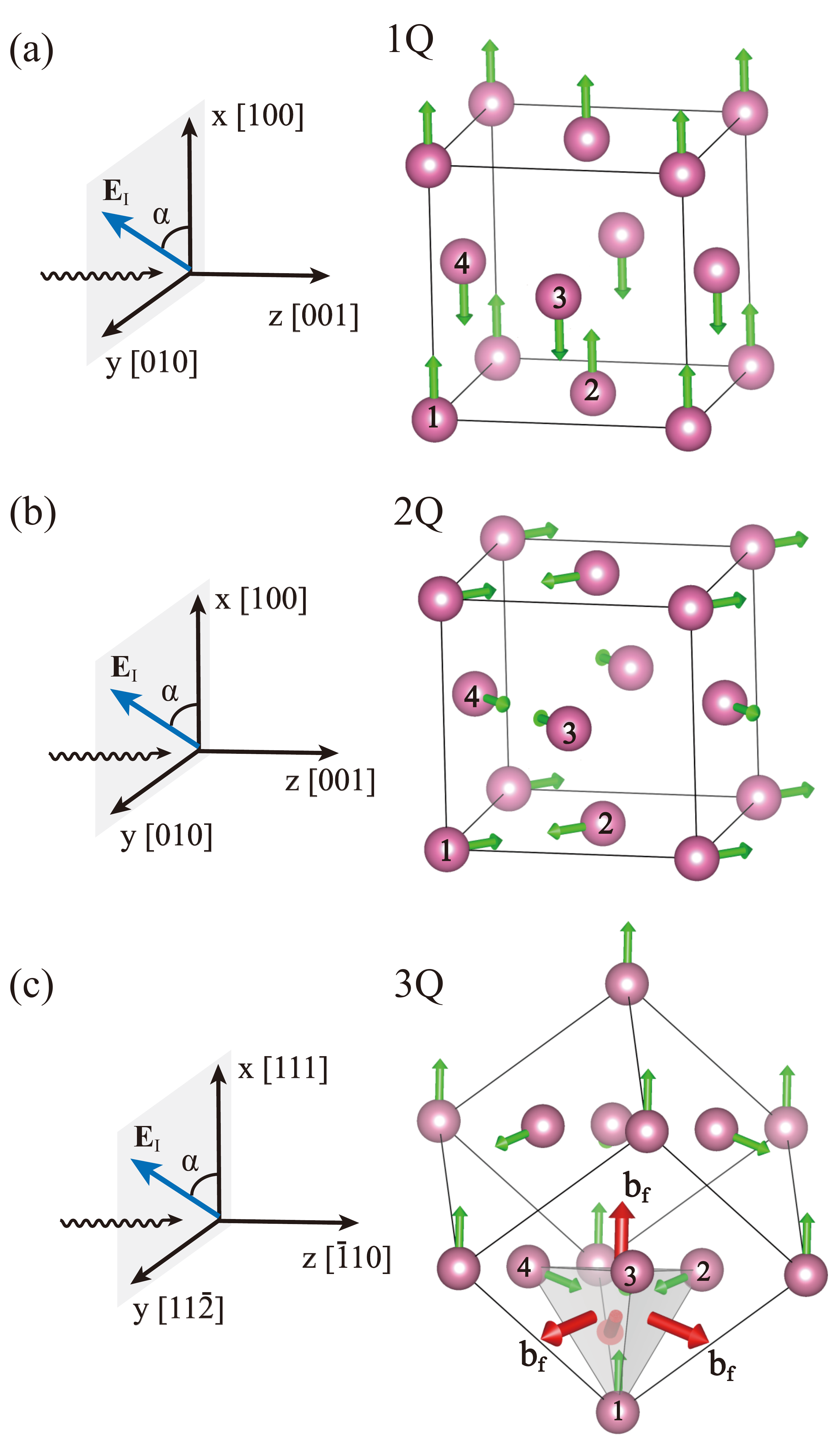}
	\caption{The 1Q (a), 2Q (b), and 3Q (c) states of $\gamma$-Fe$_{x}$Mn$_{1-x}$ as well as the optical geometries for calculating the magneto-optical Voigt and Sch\"{a}fer-Hubert effects.  The incident light propagates along the $z$-axis, and its electric field lies on the $xy$-plane with a angle of $\alpha=45^\circ$ away from the $x$-axis.  For 1Q state, the magnetization direction (i.e., the N\'{e}el vector $\mathbf{N}$) is along the $x$-axis; for 2Q state, the ``effective'' magnetization direction (i.e., $\mathbf{N}_{y}$, see the main text for details) is along the $y$-axis; for the 3Q state strained along the [111] direction, the fictitious magnetic field $\mathbf{B}=\sum_{f=1}^{4} \mathbf{b}_{f}$ is along the $x$-axis, where $\mathbf{b}_{f}$ is the fictitious magnetic flux on each face of the tetrahedron formed by four magnetic atoms in the unit cell.}
	\label{fig:crystal}
\end{figure}

The electronic structure calculations of $\gamma$-Fe$_{x}$Mn$_{1-x}$ alloy are performed using the full-potential linearized augmented-plane-wave code FLEUR~\cite{Fleur}.  The exchange-correlation effect is treated by the generalized gradient approximation with the Perdew-Burke-Ernzerhof (GGA-PBE) parameterization~\cite{Perdew1996}.  To describe disordered alloys, the virtual crystal approximation is used by adapting the nuclear numbers under conservation of charge neutrality.  The lattice constant of unstrained fcc $\gamma$-Fe$_{x}$Mn$_{1-x}$ is 6.86 $a_{0}$ (3.63 \r{A})~\cite{Hanke2017} and the muffin-tin radii of Fe and Mn atoms are chosen to be 2.29 $a_{0}$ ($a_{0}$ is Bohr radius).  A compressive or tensile strain is applied to explore how the strain influences the MOVE and MOSHE.  The strain is quantified by the distortion ratio $\delta = l'/l$,  where $l'$ and $l$ are the interatomic distances along the distorted direction in strained and unstrained structures, respectively.  The constant volume approximation is adopted on account of the Poisson effect.  The plane-wave cutoff is chosen to be 3.8 $a_{0}^{-1}$ and the Brillouin zone is sampled using a $k$-points mesh of $12\times12\times12$.  Spin-orbit coupling is not included in our calculations.  After obtaining electronic ground states, a total of 72 maximally-localized Wannier functions are constructed by projecting the $s$-, $p$-, and $d$-orbitals on four magnetic atoms in the unit cell, using \textsc{wannier90} package~\cite{Pizzi2020}.

\section{Results and Discussion} \label{sec:Results}

By varying the alloying ratio $x$,  $\gamma$-Fe$_{x}$Mn$_{1-x}$ hosts three different antiferromagnetic orders (Fig.~\ref{fig:crystal})~\cite{Kouvel1963,Endoh1971,Kubler1988,Schulthess1999,Sakuma2000}, including the collinear 1Q ($x<0.4$) and 2Q states ($x>0.8$) as well as the noncoplanar 3Q state ($0.4<x<0.8$).  In this work, we investigate the MOVE and MOSHE for 1Q, 2Q, and 3Q states by typically choosing $x$ = 0.2, 0.9, and 0.5, respectively.  In Sec.~\ref{sec:1Q}, we first discuss the conventional MOVE and MOSHE that appear in 1Q and 2Q states.  Following in Sec.~\ref{sec:3Q}, we introduce the main discovery of this work --- the topological MOVE and MOSHE, which only exists in 3Q state with a noncoplanar spin texture.

\subsection{Conventional MOVE and MOSHE in 1Q and 2Q states} \label{sec:1Q}

\begin{figure}
	\centering
	\includegraphics[width=\columnwidth]{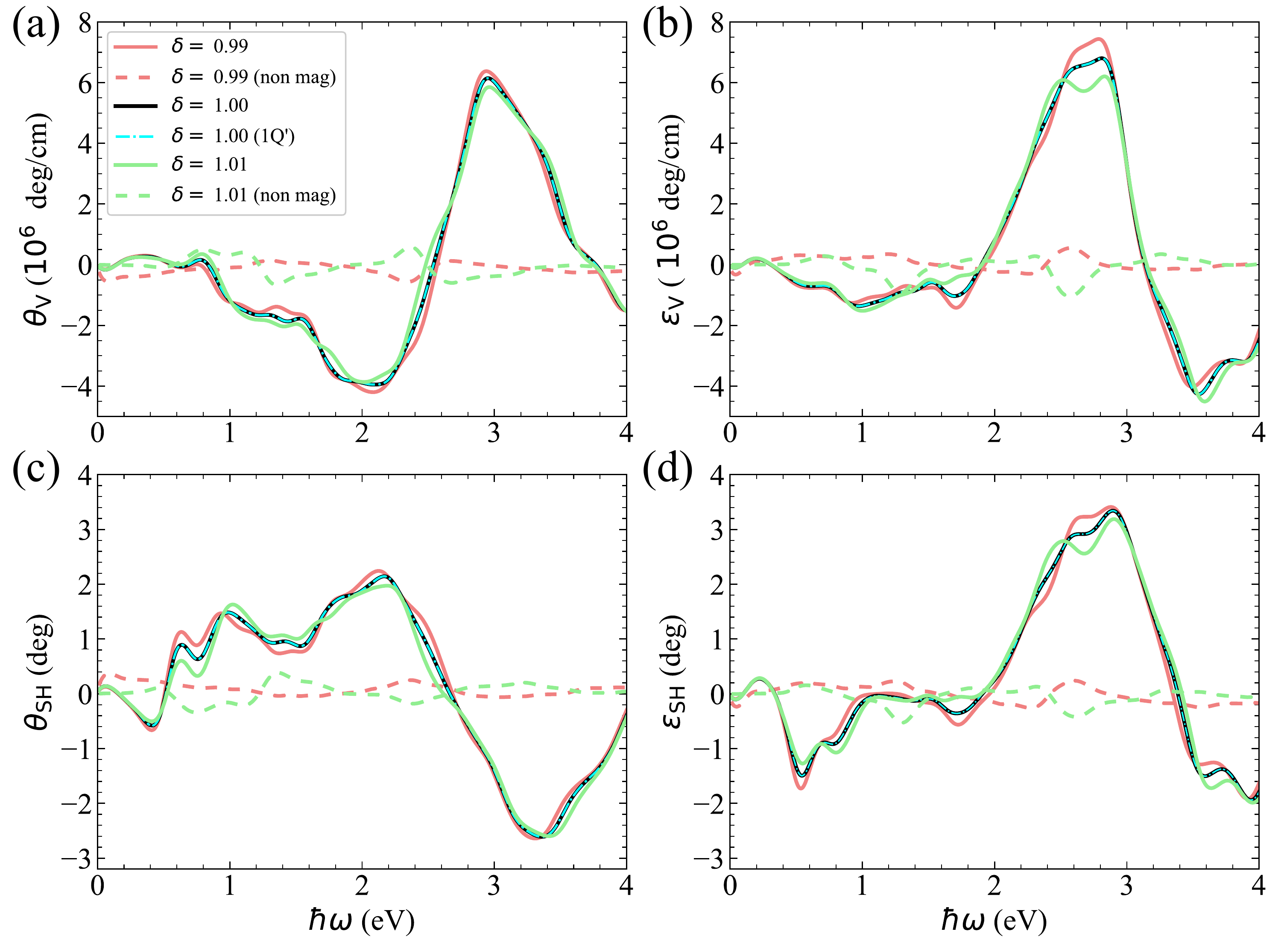}
	\caption{The Voigt and Sch\"{a}fer-Hubert rotation angles ($\theta_\textnormal{V}$ and $\theta_\textnormal{SH}$) and ellipticities ($\varepsilon_\textnormal{V}$ and $\varepsilon_\textnormal{SH}$) for the 1Q state of $\gamma$-Fe$_{x}$Mn$_{1-x}$ with $x$ = 0.2.  The compressive ($\delta$ = 0.99) and tensile ($\delta$ = 1.01) strains are applied along the [100] direction.  The 1Q$^\prime$ state (cyan dot-dashed lines) is the time-reversal counterpart of 1Q state. The natural linear birefringence (green and pink dashed lines) are calculated by removing the 1Q magnetic order on top of the two strained structures.}
	\label{fig:1Q}
\end{figure}

The 1Q state is a collinear AFM with the N\'{e}el vector, defined by $\mathbf{N}=\frac{1}{4}(\mathbf{m}_{1}+\mathbf{m}_{2}-\mathbf{m}_{3}-\mathbf{m}_{4})$ where $\mathbf{m}_{1-4}$ are spin magnetic moments on four magnetic atoms, pointing to the $x$-axis (Fig.~\ref{fig:crystal}(a)).  We first consider the perfect fcc lattice without any distortion.  The magnetic space and point groups computed by \textsc{isotropy} code~\cite{isotropy} are P$_\textnormal{I}$4/mnc (BNS setting) and 4/mmm1$^\prime$, respectively.  The symmetry-imposed shape of permittivity tensor can be identified by the linear response symmetry code \textsc{symmetr}~\cite{Zelezny2017a,Zelezny2018a},
\begin{equation}\label{eq:permittivity_1Q}
\epsilon^{\textnormal{1Q}(\textnormal{2Q})}=
	\left(\begin{array}{ccc}
		\epsilon_{xx} & 0 & 0 \\
		0 & \epsilon_{yy} & 0 \\
		0 & 0 & \epsilon_{yy}
	\end{array}\right).
\end{equation}
One can see that all the off-diagonal terms are zero and there are two independent diagonal terms, which can be simply interpreted below.  Since the time-reversal symmetry $\mathcal{T}$ is broken in antiferromagnetic 1Q state, only the antisymmetric part of off-diagonal terms ($\epsilon^{(1)}_{\mu\nu}$) is potentially zero, while the symmetric parts are always zero ($\epsilon^{(0)}_{\mu\nu}=\epsilon^{(2)}_{\mu\nu}=0$).  In addition, the magnetic point group 4/mmm1$^\prime$ contains the space-time inversion symmetry $\mathcal{PT}$, which forces $\epsilon^{(1)}_{\mu\nu}=0$.  Hence, all the off-diagonal terms have to be vanished.  For the diagonal terms, the magnetism-independent parts equal to each other in the fcc lattice, i.e., $\epsilon^{(0)}_{xx}=\epsilon^{(0)}_{yy}=\epsilon^{(0)}_{zz}$, while magnetism-dependent parts should be $\epsilon^{(2)}_{xx}\neq\epsilon^{(2)}_{yy}=\epsilon^{(2)}_{zz}$ due to the collinear antiferromagnetic order along the $x$-axis.  Accordingly, two independent diagonal terms, $\epsilon_{xx}\neq\epsilon_{yy}=\epsilon_{zz}$, appear in Eq.~\eqref{eq:permittivity_1Q}, which eventually results in the appearance of the conventional MOVE and MOSHE (refer to Eqs.~\eqref{eq:n_p-n_p},~\eqref{eq:Voigt}, and~\eqref{eq:SH}).  

Figure~\ref{fig:1Q} plots the conventional MOVE and MOSHE for the 1Q state of $\gamma$-Fe$_{x}$Mn$_{1-x}$ with $x$ = 0.2.  In the unstrained case ($\delta= 1.0$), $\theta_{\textnormal{V}}$ reaches up to $-4.0 \times 10^6 $ deg/cm at 2.1 eV  and $6.1 \times 10^6  $ deg/cm at 2.9 eV (Fig.~\ref{fig:1Q}(a)).  The spectral structure of MOSHE is similar to that of MOVE, except for a minus sign for the rotation angles (Figs.~\ref{fig:1Q}(a) and~\ref{fig:1Q}(c)).  For example, the positive and negative maximums of $\theta_{\textnormal{SH}}$ angle appear to be 2.1 deg at 2.2 eV and -2.6 deg at 3.3 eV (Fig.~\ref{fig:1Q}(c)).  The $\theta_{\textnormal{V}}$ and $\theta_{\textnormal{SH}}$ of 1Q state are larger than that of other collinear AFMs, such as CuMnAs ($\theta_\textnormal{V} \sim 2.3\times10^4 $ deg/cm)~\cite{Saidl2017}, CoO thin film ($\theta_\textnormal{SH} \sim  0.168 $ deg)~\cite{J-Xu2020}.  If we apply the time-reversal symmetry for 1Q state (named 1Q$^\prime$ state), all the spin magnetic moments ($\mathbf{m}_{1-4}$) change their signs such that the N\'{e}el vector $\mathbf{N}$ reverses its direction.  Nevertheless, $\theta_{\textnormal{V}}$ and $\theta_{\textnormal{SH}}$ change nothing (see cyan dot-dashed lines).  It indicates that the observed MOVE and MOSHE are essentially even in $\mathbf{N}$, differing from the linear MOKE and MOFE that are odd in $\mathbf{N}$.

In practice, a small tetragonal distortion along the direction of $\mathbf{N}$ possibly occurs in 1Q state according to the previous works~\cite{Ekholm2011}, similarly to pure $\gamma$-Fe~\cite{Ehrhart1980} and pure $\gamma$-Mn~\cite{Oguchi1984}.  Here, 1\% compressive ($\delta=0.99$) and tensile ($\delta=1.01$) strains along the $x$-axis (i.e., [100] direction) are taken into account.  The magnetic space and point groups remain unchanged, that is, P$_\textnormal{I}$4/mnc (BNS setting) and 4/mmm1$^\prime$, respectively.  Since the $\mathcal{PT}$ symmetry is reserved under the strains, all the off-diagonal terms of permittivity tensor are still zero.  The shape of permittivity tensor thus keeps invariant, see Eq.~\eqref{eq:permittivity_1Q}.  The strains only affect the magnetism-independent diagonal terms, leading to $\epsilon^{(0)}_{xx}\neq\epsilon^{(0)}_{yy}=\epsilon^{(0)}_{zz}$, which induces the NLB.  Figure~\ref{fig:1Q} clearly shows that the influence of strains on the MOVE and MOSHE for 1Q state is rather weak (comparing the lines for $\delta=0.99$, $1.0$, and $1.01$).   The difference between the strained and unstrained results is just the contribution from NLB, which can also be calculated by directly plugging $\epsilon^{(0)}_{xx}$ and $\epsilon^{(0)}_{yy}$ into Eqs.~\eqref{eq:n_pn_p} and~\eqref{eq:n_p-n_p} (see green and pink dashed lines in Fig.~\ref{fig:1Q}).  The NLB under compressive and tensile strains along the $x$-axis have almost the same spectral structure but differ by a minus sign, which reflects the dominated absorption of the linearly polarized light along either $x$-axis or $y$-axis.  Nevertheless, the NLB for both compressive and tensile cases are considerably weak, at least one order of magnitude smaller than the MOVE and MOSHE.

\begin{figure}
	\centering
	\includegraphics[width=\columnwidth]{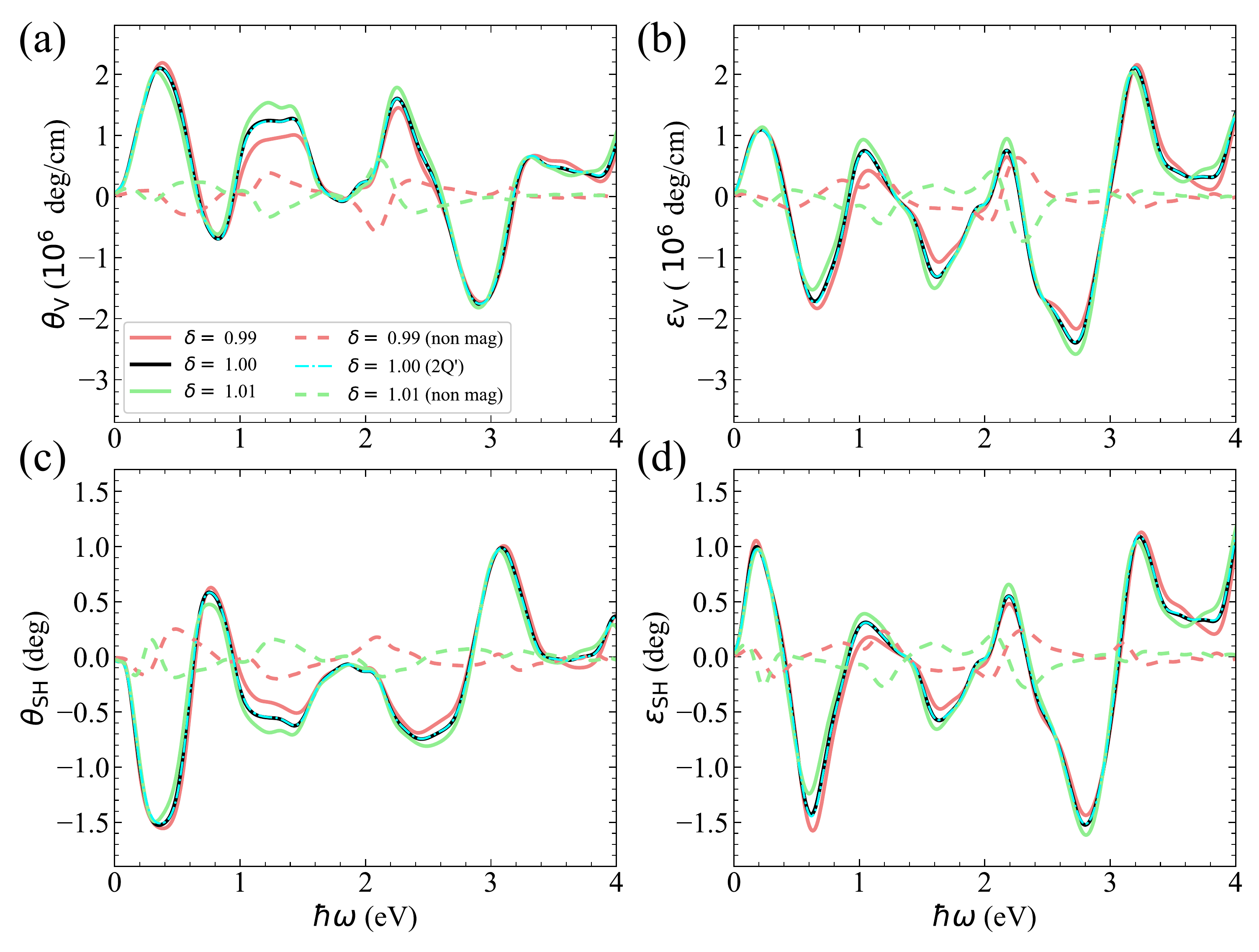}
	\caption{The Voigt and Sch\"{a}fer-Hubert rotation angles ($\theta_\textnormal{V}$ and $\theta_\textnormal{SH}$) and ellipticities ($\varepsilon_\textnormal{V}$ and $\varepsilon_\textnormal{SH}$) for the 2Q state of $\gamma$-Fe$_{x}$Mn$_{1-x}$ with $x$ = 0.9.  All the labels are the same as the 1Q state presented in Fig.~\ref{fig:1Q}.}
	\label{fig:2Q}
\end{figure}

Next, we turn to the MOVE and MOSHE for 2Q state, which is also a collinear AFM since both the vector and scalar spin chiralities are zero ($\boldsymbol{\kappa}=0$ and $\chi=0$).  The 2Q state has two collinear antiferromagnetic orders along the [0$\bar{1}$1] and [011] directions, which are orthogonal to each other.  The calculated magnetic space and point groups of 2Q state are P$_\textnormal{C}$4$_2$/nnm (BNS setting) and 4/mmm1$^\prime$, respectively.  The shape of the permittivity tensor for 2Q state is identical to 1Q state, as given in Eq.~\eqref{eq:permittivity_1Q}.  The vanishing off-diagonal terms suggest that the first-order MOKE and MOFE must be absent, while the unequal diagonal terms imply the existence of second-order MOVE and MOSHE.  However, at first glance, the magnetization direction of 2Q state does not satisfy the standard Voigt geometry (Fig.~\ref{fig:crystal}(b) and Fig.~\ref{fig:model}).  Here we decompose the spin magnetic moment of each atom to the $y$- and $z$-axes.  The system is still a compensated AFM as the net magnetization along both the $y$- and $z$-axes are zero.  The collinear antiferromagnetic order along the $z$-axis,  $\mathbf{N}_{z}=\frac{1}{4}(\mathbf{m}_{1,z}-\mathbf{m}_{2,z}-\mathbf{m}_{3,z}+\mathbf{m}_{4,z})$, plays no role in the first- and second-order magneto-optical effects, while the one along the $y$-axis, $\mathbf{N}_{y}=\frac{1}{4}(-\mathbf{m}_{1,y}+\mathbf{m}_{2,y}-\mathbf{m}_{3,y}+\mathbf{m}_{4,y})$, activates the second-order MOVE and MOSHE.  The angle $\alpha$ between the electric field direction of incident light ($\mathbf{E}_\textnormal{I}$) and the ``effective" magnetization direction ($\mathbf{N}_{y}$) is 45$^\circ$ (Fig.~\ref{fig:crystal}(b)), giving rise to the maximal values of $\theta_{\textnormal{V}}$ and $\theta_{\textnormal{SH}}$.

Figure~\ref{fig:2Q} plots the conventional MOVE and MOSHE for the 2Q state of $\gamma$-Fe$_{x}$Mn$_{1-x}$ with $x$ = 0.9.  Similarly to 1Q state, the spectral structures of MOVE and MOSHE resemble each other, only a minus sign differs for the rotation angles (Figs.~\ref{fig:2Q}(a) and~\ref{fig:2Q}(c)).  On the other hand, in contrast to 1Q state, $\theta_{\textnormal{V(SH)}}$ and $\varepsilon_{\textnormal{V(SH)}}$ oscillate frequently as a function of photon energy.  The positive and negative maximums of $\theta_{\textnormal{V}}$ ($\theta_{\textnormal{SH}}$) are $2.1\times 10^6$ deg/cm at 0.4 eV and $-1.7\times 10^6$ deg/cm at 2.9 eV ($1.0^{\circ}$ at 3.1 eV and $-1.5^{\circ}$ at 0.4 eV), respectively.  The MOVE and MOSHE of 2Q state are overall smaller than that of 1Q state.  It can be understood from the fact that the effective antiferromagnetic order of 2Q state (along the $y$-axis) is weaker than the true antiferromagnetic order of 1Q state (along the $x$-axis) because the spin magnetic moments of 2Q state projected onto the plane perpendicular to the incident light ($xy$-plane) are actually reduced.  This is very similar to the case of CuMnAs when rotating $\mathbf{N}$ around an axis perpendicular to the direction of the spin magnetic moments~\cite{Saidl2017}.  If the time-reversal symmetry is applied on 2Q state (named 2Q$^\prime$ state), all the moments will be reversed such that the effective antiferromagnetic order $\mathbf{N}_{y}$ changes its direction.  By comparing to 2Q state, the unchanged MOVE and MOSHE of 2Q$^\prime$ state demonstrate that the magneto-optical effects being even in $\mathbf{N}_{y}$ are indeed second-order.  The effect of strain on the MOVE and MOSHE for 2Q state is also considered by applying a 1\% compressive ($\delta=0.99$) or tensile  ($\delta=1.01$) strain along the $x$-axis.  The resultant NLB is significantly smaller than the magnetism-induced MOVE and MOSHE, similarly to 1Q state.  The strain-insensitive character is also similar to the optical linear dichroism in two-dimensional zigzag-AFM FePS$_{3}$~\cite{Q-Zhang2021}.

\begin{figure*}
	\centering
	\includegraphics[width=2\columnwidth]{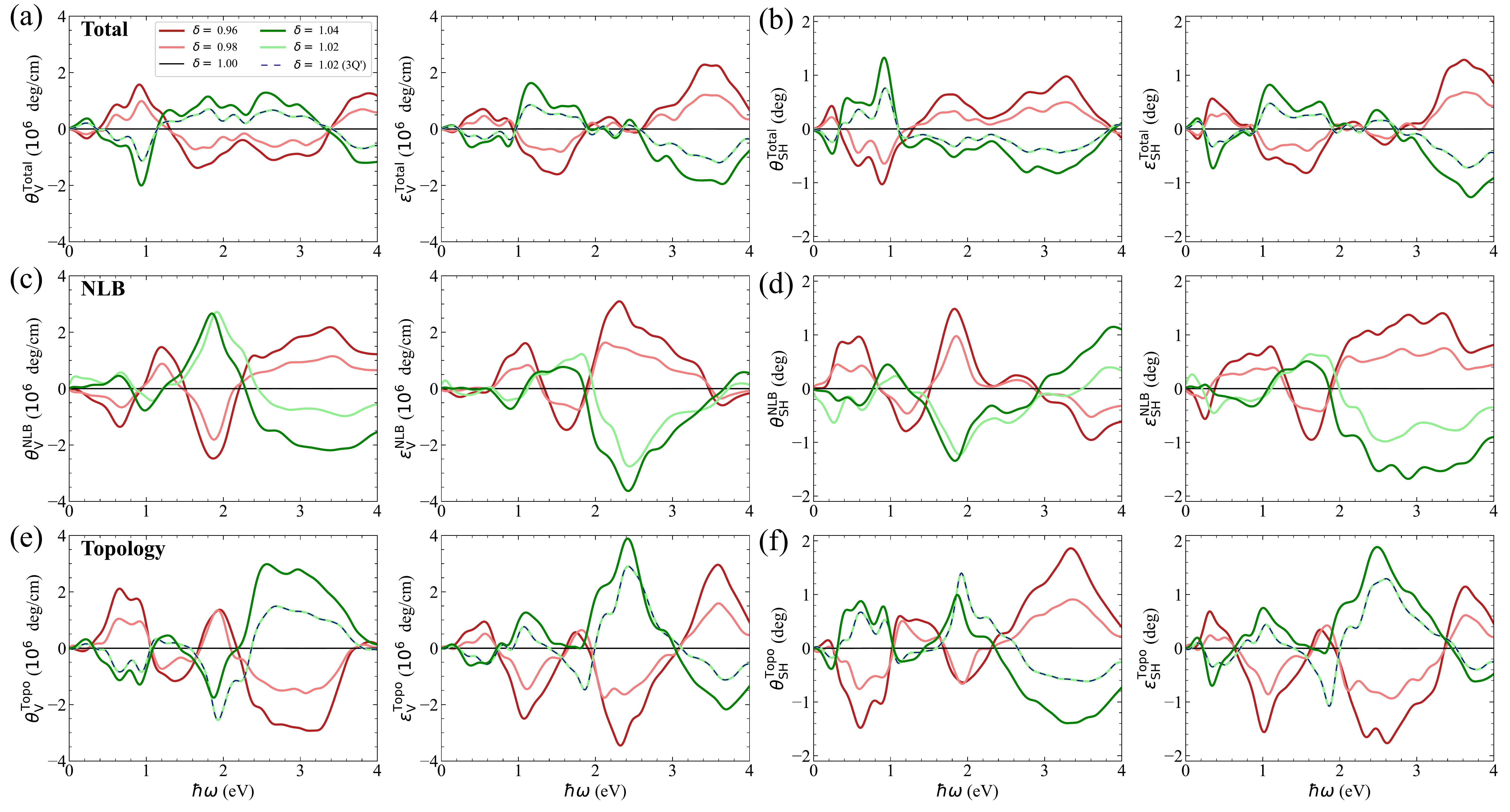}
	\caption{The Voigt and Sch\"{a}fer-Hubert rotation angles ($\theta_\textnormal{V}$ and $\theta_\textnormal{SH}$) and ellipticities ($\varepsilon_\textnormal{V}$ and $\varepsilon_\textnormal{SH}$) for the strained 3Q state of $\gamma$-Fe$_{x}$Mn$_{1-x}$ with $x$ = 0.5.  (a)(b) The total angles contributed from both crystal anisotropy and scalar spin chirality.  (c)(d) The contribution solely from crystal anisotropy, that is, natural linear birefringence.  (e)(f) The contribution solely from scalar spin chirality, that is, topological Voigt and Sch\"{a}fer-Hubert effects.  The compressive ($\delta$ = 0.96, 0.98) and tensile ($\delta$ = 1.02, 1.04) strains are applied along the [111] direction.  The 3Q$^\prime$ state (blue dashed lines) is the time-reversal counterpart of 3Q state.}
	\label{fig:3Q}
\end{figure*}

\subsection{Topological MOVE and MOSHE in 3Q state}\label{sec:3Q}

The 3Q state is a fully compensated noncoplanar AFM, in which all spin magnetic moments point to the center of the tetrahedron formed by four magnetic atoms in the fcc lattice (Fig.~\ref{fig:crystal}(c)).   Considering one face of the tetrahedron, the three noncoplanar spins generate a nonzero scalar spin chirality, $\chi_{ijk} = \mathbf{S}_{i}\cdot\left(\mathbf{S}_{j}\times\mathbf{S}_{k}\right) \neq 0$, which in turn provides a fictitious magnetic flux, $\mathbf{b}_{f}\propto t_{3}\chi_{ijk}\mathbf{\hat{n}}$ (here, $t_{3}=t_{ij}t_{jk}t_{ki}$ is the successive transfer integral along the path $i\rightarrow j \rightarrow k \rightarrow i$ and $\mathbf{\hat{n}}$ is the surface normal vector)~\cite{WX-Feng2020}.  The total fictitious magnetic field in the unit cell is the vector sum of the magnetic fluxes on the four faces of the tetrahedron, i.e., $\mathbf{B}=\sum_{f=1}^{4}\mathbf{b}_{f}$.  It is clear that $\mathbf{B}$ is zero for the unstrained fcc lattice because the four fluxes cancel each other exactly  (Fig.~\ref{fig:crystal}(c)).  Accordingly, both the first- and second-order magneto-optical effects are not active in the unstrained 3Q state.  This can be further understood from the symmetry point of view.  For the unstrained 3Q state, the magnetic space and point groups are $Pn\bar{3}m^{\prime}$ and $m\bar{3}m^{\prime}$, respectively.  The resultant permittivity tensor shows vanishing off-diagonal terms and equivalent diagonal terms,
\begin{equation}\label{eq:permittivity_3Q}
	\epsilon^{\textnormal{3Q}}=
	\left(\begin{array}{ccc}
		\epsilon_{xx} & 0 & 0 \\
		0 & \epsilon_{xx} & 0 \\
		0 & 0 & \epsilon_{xx}
	\end{array}\right),
\end{equation}
which forbids both the first- and second-order magneto-optical effects.

The scalar spin chirality in 3Q state plays its role once a strain is applied on the fcc lattice.  For example, the topological orbital magnetization and topological Hall effect, originating from scalar spin chirality, were reported in strained $\gamma$-Fe$_{x}$Mn$_{1-x}$ with the 3Q spin texture~\cite{Hanke2017,Shiomi2018}.  Moreover, the first-order topological magneto-optical effects (by taking MOKE and MOFE as prototypes) and their quantization were predicted in our previous work~\cite{WX-Feng2020}.  One can rationally speculate that the second-order topological magneto-optical effects, i.e., topological MOVE and MOSHE, exist also in the strained 3Q state of $\gamma$-Fe$_{x}$Mn$_{1-x}$.  To confirm this idea, a uniaxial strain is applied along the [111] direction such that a nonzero fictitious magnetic field, $\mathbf{B}=B\mathbf{\hat{n}}_{[111]}$ with $B\neq0$, emerges.  Further assuming that the light is incident along the [$\bar{1}$10] direction, and its electric field is 45$^\circ$ away from the [111] direction.  The optical geometry for calculating the topological MOVE and MOSHE of the strained 3Q state is schematically displayed in Fig.~\ref{fig:crystal}(c).  The magnetic space and point groups for the strained 3Q state are $R\bar{3}m^{\prime}$ and $\bar{3}1m^{\prime}$, respectively, giving rise to the permittivity tensor,
\begin{equation}\label{eq:permittivity_3Q_strained}
	\epsilon^{\textnormal{3Q,strained}}=
	\left(\begin{array}{ccc}
		\epsilon_{xx} & 0 & 0 \\
		0 & \epsilon_{yy} & \epsilon_{yz} \\
		0 & -\epsilon_{yz} & \epsilon_{yy}
	\end{array}\right).
\end{equation}
 Due to the unequal diagonal terms ($\epsilon_{xx}\neq\epsilon_{yy}$) and nonzero off-diagonal terms ($\epsilon_{yz}\neq0$), the $\theta_{\textnormal{V}}$ and $\theta_{\textnormal{SH}}$ are certainly nonvanishing for the strained 3Q state (refer to Eqs.~\eqref{eq:n_p-n_p},~\eqref{eq:Voigt}, and~\eqref{eq:SH}).  However, the topological component originating from scalar spin chirality is mixed to the NLB that is due to crystal anisotropy.  To obtain the topological MOVE and MOSHE, the NLB has to be separated from total birefringence.
 
 Figure~\ref{fig:3Q} illustrates the Voigt and Sch\"{a}fer-Hubert spectra for the 3Q state of $\gamma$-Fe$_{x}$Mn$_{1-x}$ with $x$ = 0.5.  The total angles ($\theta_{\textnormal{V}}^{\textnormal{Total}}$, $\varepsilon_{\textnormal{V}}^{\textnormal{Total}}$,  $\theta_{\textnormal{SH}}^{\textnormal{Total}}$, and $\varepsilon_{\textnormal{SH}}^{\textnormal{Total}}$) contributed from both crystal anisotropy and scalar spin chirality are shown in Figs.~\ref{fig:3Q}(a) and~\ref{fig:3Q}(b).  In the unstrained case ($\delta$ = 1.0), the Voigt/Sch\"{a}fer-Hubert rotation angles and ellipticities are definitely zero, which is consistent with the above symmetry analysis, referring to Eq.~\eqref{eq:permittivity_3Q}.  Once a strain, for example, along the [111] direction,  is applied ($\delta\neq$ 1.0), the rotation angles and ellipticities turn to be nonzero and are roughly proportional to the strain.  Moreover, the direction of the angles reverses if the applied strain changes from tension ($\delta>$ 1.0) to compression ($\delta<$ 1.0) and vice versa.  As mentioned above, the strain brings crystal anisotropy and finite fictitious magnetic field simultaneously in the fcc lattice, and therefore, the observed total Voigt and Sch\"{a}fer-Hubert angles should be a superposition of the NLB and topological MOVE/MOSHE.
 
 If we remove the 3Q magnetic order on top of the strained fcc lattices, the off-diagonal terms of the permittivity tensor should be zero ($\epsilon_{yz}=\epsilon_{zy}=0$), while the diagonal terms retain only the magnetism-independent parts ($\epsilon_{xx}^{(0)}\neq\epsilon_{yy}^{(0)}\neq 0$).  In such a case, the calculated Voigt and Sch\"{a}fer-Hubert angles can be completely ascribed to the NLB ($\theta_{\textnormal{V}}^{\textnormal{NLB}}$, $\varepsilon_{\textnormal{V}}^{\textnormal{NLB}}$,  $\theta_{\textnormal{SH}}^{\textnormal{NLB}}$, and $\varepsilon_{\textnormal{SH}}^{\textnormal{NLB}}$) that is due to crystal anisotropy, as shown in Figs.~\ref{fig:3Q}(c) and~\ref{fig:3Q}(d).  The NLB generated by the strain along the [111] direction is three to four times larger than that along the [100] direction (comparing Figs.~\ref{fig:3Q}(c,d) with Figs.~\ref{fig:1Q} and~\ref{fig:2Q}).  On the other hand, the topological components of Voigt and Sch\"{a}fer-Hubert angles ($\theta_{\textnormal{V}}^{\textnormal{Topo}}$, $\varepsilon_{\textnormal{V}}^{\textnormal{Topo}}$,  $\theta_{\textnormal{SH}}^{\textnormal{Topo}}$, and $\varepsilon_{\textnormal{SH}}^{\textnormal{Topo}}$), i.e., topological MOVE and MOSHE that root in scalar spin chirality, can be obtained by solely taking into account the magnetism-dependent terms of permittivity tensor (including $\epsilon_{xx}^{(2)}$, $\epsilon_{yy}^{(2)}$, and $\epsilon_{yz}^{(1)}$), as shown in Figs.~\ref{fig:3Q}(e) and~\ref{fig:3Q}(f).  One should note that the magnitudes of topological MOVE and MOSHE are in the same order of the NLB.  In addition, the Voigt and Sch\"{a}fer-Hubert angles originated from crystal anisotropy and scalar spin chirality own opposite signs (Figs.~\ref{fig:3Q}(c-f)), reflecting the contrary deflection of the polarization plane when a linearly polarized light goes through the strained 3Q state.
 
By applying the time-reversal operation $\mathcal{T}$, all the spin magnetic moments of 3Q state are inverted, giving the 3Q$^{\prime}$ state, in which the fictitious magnetic fluxes $\mathbf{b}_{f}$ on each face of the tetrahedron reverse their directions.  As a result, the total fictitious magnetic filed $\mathbf{B}$ in the unit cell of 3Q$^{\prime}$ state also reverses its direction.  One can see that the topological MOVE and MOSHE for 3Q and 3Q$^{\prime}$ states are fully identical to each other, as plotted in Figs.~\ref{fig:3Q}(e) and~\ref{fig:3Q}(f), by taking the strain $\delta$ = 1.02 as an example (others are not shown).  It demonstrates clearly that  the topological MOVE and MOSHE belong to the second-order magneto-optical effects as they are even in $\mathbf{b}_{f}$ or even in $\mathcal{T}$ equivalently.  Here, it is not safe to say that the topological MOVE and MOSHE are even in the fictitious magnetic filed $\mathbf{B}$ since the direction of $\mathbf{B}$ can also be inverted by applying the strain from tension to compression and vice versa, while during this process the signs of topological MOVE and MOSHE follow with $\mathbf{B}$.  The strain does not change the direction of $\mathbf{b}_{f}$ but creates an imbalance among the $\mathbf{b}_{f}$ on the four faces of the tetrahedron, which in turn reverses the direction of $\mathbf{B}$.  Hence, such a change in the direction of $\mathbf{B}$ by strain is irrelevant to time-reversal symmetry.  While the advantage is that the strain provides a practical way to control the signs of topological MOVE and MOSHE, which can not be realized in conventional MOVE and MOSHE.  Finally, one can also find that the total Voigt and Sch\"{a}fer-Hubert angles of 3Q and 3Q$^{\prime}$ states are in a complete agreement, as shown in Figs.~\ref{fig:3Q}(a) and~\ref{fig:3Q}(b), because $\mathcal{T}$ plays no role in the NLB.

\begin{figure}
	\centering
	\includegraphics[width=1.0\columnwidth]{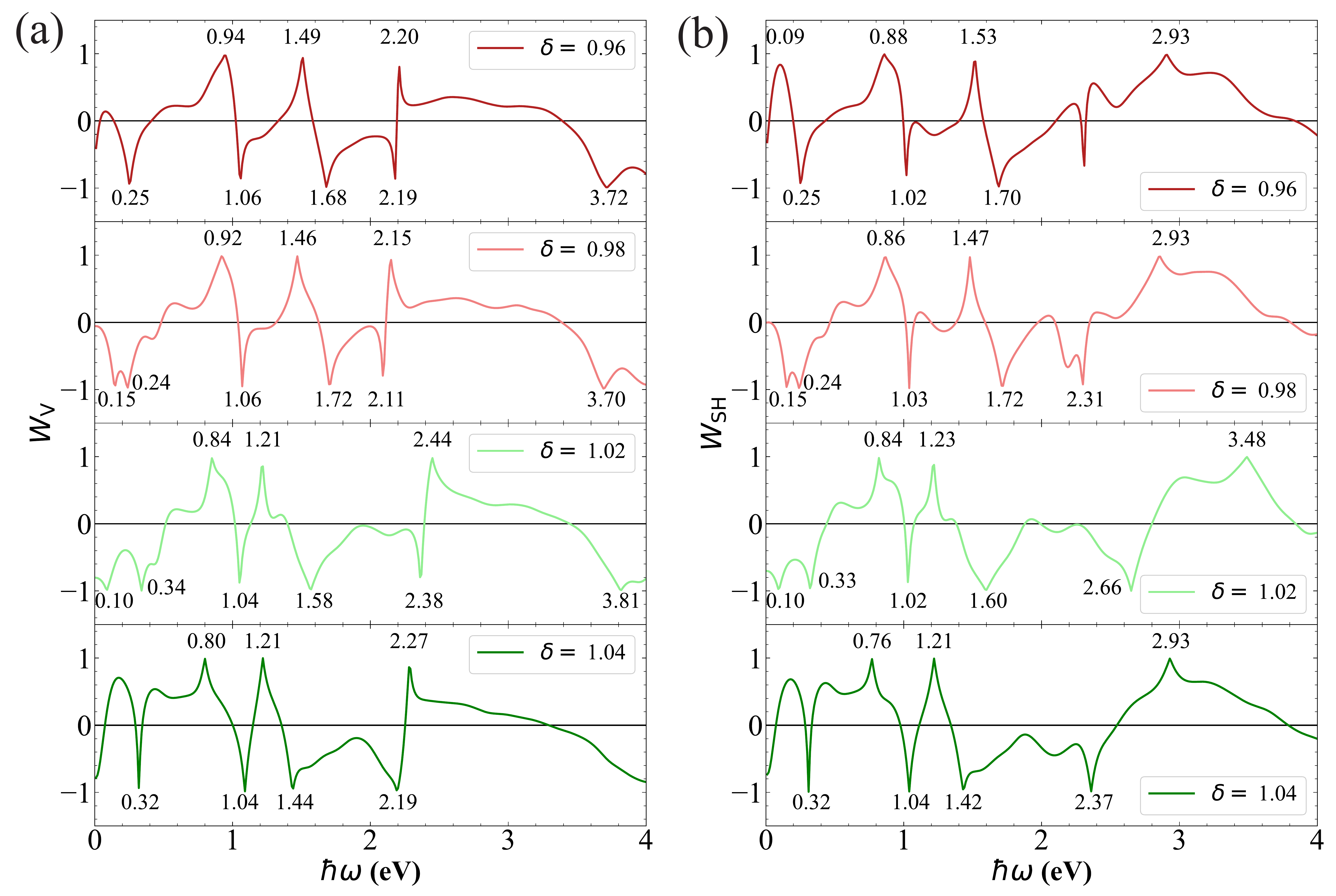}
	\caption{The relative weight of topological component and natural linear birefringence for total Voigt and Sch\"{a}fer-Hubert rotation angles ($W_{\textnormal{V}}$ and $W_{\textnormal{SH}}$) of the strained 3Q state of $\gamma$-Fe$_{x}$Mn$_{1-x}$ ($x$ = 0.5 and $\delta$ = 0.96, 0.98, 1.02, 1.04).  The photon energies are marked when $W_{\textnormal{V}}$ and $W_{\textnormal{SH}}$ approach to $\pm1$.}
	\label{fig:3Q_weight}
\end{figure}

In the above theoretical calculations, total Voigt and Sch\"{a}fer-Hubert rotation angles ($\theta_{\textnormal{V}}^{\textnormal{Total}}$ and $\theta_{\textnormal{SH}}^{\textnormal{Total}}$) contributed from topological origin ($\theta_{\textnormal{V}}^{\textnormal{Topo}}$ and $\theta_{\textnormal{SH}}^{\textnormal{Topo}}$) and NLB ($\theta_{\textnormal{V}}^{\textnormal{NLB}}$ and $\theta_{\textnormal{SH}}^{\textnormal{NLB}}$) have been explicitly separated.  On the other hand, for experiments, the NLB part can also be captured by using, for example, the ultrafast pump-probe technique, which can induce demagnetization in AFMs~\cite{HC-Zhao2021,Saidl2017}.  To describe the relative weight of topological origin and NLB quantitatively, we define such a quantity,
\begin{equation}\label{eq:3Q_weigt}
W_{\textnormal{V(SH)}} = \frac{|\theta_{\textnormal{V(SH)}}^{\textnormal{Topo}}| - |\theta_{\textnormal{V(SH)}}^{\textnormal{NLB}}|}{|\theta_{\textnormal{V(SH)}}^{\textnormal{Topo}}| + |\theta_{\textnormal{V(SH)}}^{\textnormal{NLB}}|},
\end{equation}
where $W_{\textnormal{V}}$ and $W_{\textnormal{SH}}$ approaching to 1 (-1) means that the topological (NLB) component dominates.  Figure~\ref{fig:3Q_weight} depicts the $W_{\textnormal{V}}$ and $W_{\textnormal{SH}}$ for the 3Q state of $\gamma$-Fe$_{x}$Mn$_{1-x}$ ($x$ = 0.5) under four different strains ($\delta$ = 0.96, 0.98, 1.02, and 1.04).  Taking $W_{\textnormal{V}}$ at $\delta$ = 0.96 as an example, the first positive peak appears at 0.94 eV, at where $\theta_{\textnormal{V}}^{\textnormal{Total}}$, $\theta_{\textnormal{V}}^{\textnormal{Topo}}$, $\theta_{\textnormal{V}}^{\textnormal{NLB}}$ are $1.52\times10^{6}$ deg/cm, $1.53\times10^{6}$ deg/cm, and $-0.02\times10^{6}$ deg/cm, respectively (see Fig.~\ref{fig:3Q}(a,c,e), left panels); the first negative pear appears at 0.25 eV, at where $\theta_{\textnormal{V}}^{\textnormal{Total}}$, $\theta_{\textnormal{V}}^{\textnormal{Topo}}$, $\theta_{\textnormal{V}}^{\textnormal{NLB}}$ are $-0.32\times10^{6}$ deg/cm, $-0.31\times10^{6}$ deg/cm, and $-0.01\times10^{6}$ deg/cm, respectively (see Fig.~\ref{fig:3Q}(a,c,e), left panels).  It confirms that $W_{\textnormal{V}}$ and $W_{\textnormal{SH}}$ are precise fingerprints for experimentally probing the topological MOVE and MOSHE in the strained 3Q state of $\gamma$-Fe$_{x}$Mn$_{1-x}$.

\section{Summary} \label{sec:Summary}

In summary, the second-order magneto-optical effects have been investigated in the collinear 1Q and 2Q states as well as the noncoplanar 3Q state of antiferromagnetic $\gamma$-Fe$_{x}$Mn$_{1-x}$ alloy, using the first-principles calculations and group theory analysis.  The conventional Voigt and Sch\"{a}fer-Hubert effects were found in the collinear 1Q and 2Q states, just like other common antiferromagnets.  While the Voigt and Sch\"{a}fer-Hubert rotation angles emerged in the 1Q state reach up to $6.1 \times 10^6  $ deg/cm and 2.6 deg, respectively, which are much larger than that of some famous collinear antiferromagnets, e.g., CuMnAs.  On the other hand, the Voigt and Sch\"{a}fer-Hubert rotation angles emerged in the 2Q state are relatively small since the effective magnetization perpendicular to the incident light is reduced.  The natural linear birefringence originating from crystal anisotropy was observed in the strained 1Q and 2Q states, but their magnitudes are notably less than the magnetism-induced Voigt and Sch\"{a}fer-Hubert effects.  In the strained 3Q state, the total Voigt and Sch\"{a}fer-Hubert angles were contributed from the topological origin and natural linear birefringence because the strain brings scalar spin chirality and crystal anisotropy simultaneously.  The topological Voigt and Sch\"{a}fer-Hubert effects, originating from scalar spin chirality, were successfully identified.  A unique fingerprint for experimentally probing the topological Voigt and Sch\"{a}fer-Hubert effects was also proposed.  Our work not only deepens the understanding of second-order magneto-optical effects but also facilitates the applications of antiferromagnets in magneto-optical devices.

\begin{acknowledgments}
This work is supported by the National Natural Science Foundation of China (Grant Nos. 11874085, 11734003, and 12061131002), the Sino-German Mobility Programme (Grant No. M-0142), the National Key R\&D Program of China (Grant No. 2020YFA0308800), and the Science \& Technology Innovation Program of Beijing Institute of Technology (Grant No. 2021CX01020).
\end{acknowledgments}


%

\end{document}